\documentclass[aps,prb,twocolumn,groupedaddress,floatfix]{revtex4-1}

\usepackage{graphicx}
\usepackage{dcolumn}
\usepackage{bm}
\usepackage{color}

\begin{document}

\title{Thermo-electromechanical effects in  relaxed shape graphene  and  bandstructures of graphene quantum dots}
\author{Sanjay Prabhakar,$^1$  Roderick Melnik,$^{1}$ Luis L. Bonilla$^2$ and Shyam Badu$^1$ }
\affiliation{
$^1$The MS2Discovery Interdisciplinary Research Institute, M\,$^2$NeT Laboratory, Wilfrid Laurier University, Waterloo, ON, N2L 3C5 Canada\\
$^2$Gregorio Millan Institute, Universidad Carlos III de Madrid, 28911, Leganes, Spain
}
\date{September 24, 2014}

\begin{abstract}
We investigate the in-plane oscillations of the relaxed shape graphene due to externally applied tensile edge stress along both the armchair and zigzag directions.
We show that the total  elastic energy density is enhanced with temperature  for the case of applied tensile edge stress along the zigzag direction.
Thermo-electromechanical effects are treated via pseudomorphic vector potentials to analyze the influence of these coupled effects on the bandstructures of bilayer graphene quantum dots (QDs).   We report that the level crossing between ground and first excited states in the localized edge states can be achieved with the accessible values of temperature. In particular, the level crossing point extends  to higher temperatures with  decreasing values of  externally applied tensile edge stress along the armchair direction. This kind of level crossings is absent  in the states formed at the center of the graphene sheet due to the presence of three fold symmetry.
\end{abstract}



\maketitle

\section{Introduction}

Miniaturization is one of the requirements to make  electronic devices smaller and smaller. To fulfill this requirement, graphene based electronic devices might provide a breakthrough alternative for   the current state-of-the-art semiconductor industry. Graphene is a promising material to build electronic devices because of its unusual  properties due to the Dirac-like spectrum of the charge carriers.~\cite{novoselov04,novoselov05}
In addition, researchers around the world  seek  to make  next generation electronic devices from graphene  because the material possesses high charge mobility. There is an opportunity to control electronic properties of graphene-based structures
by several different techniques such as gate controlled electric fields, magnetic fields, and to engineer the electromechanical properties  via   pseudomorphic gauge fields.~\cite{neto09,shenoy08,choi10,maksym13,bao12,yan12,cadelano09,cadelano12,bao-lau09} Recently,   spin echo phenomena, followed by strong beating patterns  due to  rapid oscillations of quantum states along a graphene ribbon,  have been investigated for  realization of quantum memory at optical states in the far-infrared region for quantum information processing.~\cite{prabhakar13,belonenko12}

Graphene sheets  designed to make  next generation electronic devices consist of surfaces that are not perfectly flat. The  surfaces exhibit intrinsic microscopic roughening,  the surface normal varies by several degrees and the out-of-plane deformations reach to the nanometer scale.~\cite{gass08,shenoy08,meyer07,bonilla12,guinea10} Out-of-plane displacements  without a preferred direction induce ripples in the graphene sheet,~\cite{gibertini10,wang12,bonilla12,carpio08,sanjose11} while in-plane displacements induced  by applying suitable tensile edge stress along the armchair and zigzag directions lead to the relaxed shape  graphene.~\cite{gass08,shenoy08}
In experiments with graphene  suspended on substrate trenches, there appear much longer and taller waves (close to a micron scale) directed parallel to the applied stress.~\cite{gass08,shenoy08,cerda-mahadevan03} These long wrinkles are mechanically induced and we know that theoretically it is possible to corrugate an elastic membrane at absolute zero temperature.
The mechanical deformation of a free standing graphene sheet can be understood by applying suitable edge stress along the armchair and zigzag directions.
The  tensile forces, that  can be applied on the graphene sheet with a compressed elastic string, are schematically shown in Fig.~\ref{fig1}.
In this paper, we show that intrinsic edge stresses can have a significant influence on the morphology of graphene sheets and substantially modify the bandstructures of graphene quantum dots.
We present a model that couples the Navier equations, accounting for thermo-electromechanical effects,  to the electronic properties of graphene quantum dots.~\cite{zakharchenko09} Here we show  that the amplitude of the  induced waves due to applied tensile edge stress along the armchair and zigzag directions increases with temperature and provides the level crossing  in the localized edge states. This kind of level crossings is absent for the states formed at the center of the graphene sheet due to the presence of three fold symmetry induced by pseudomorphic strain tensor.


\section{Theoretical Model}\label{theoretical-model}

\begin{figure}
\includegraphics[width=7cm,height=5cm]{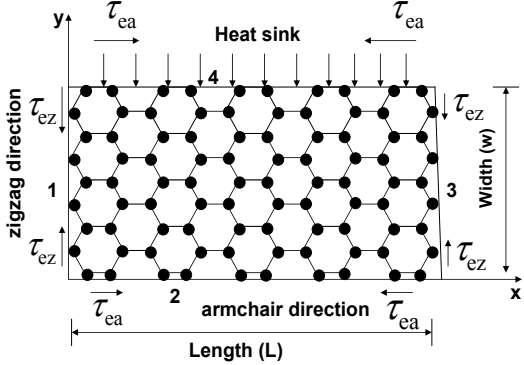}
\caption{\label{fig1}
Tensile stress is applied  along armchair and zigzag directions that induce the oscillation in strain tensor in graphene sheet. The boundary 4 is connected to the heat bath to investigate the influence of temperature on the strain tensor as well as on the bandstructures of graphene.
}
\end{figure}

The total thermoelastic energy density  associated to the strain for the two-dimensional graphene sheet can be written  as~\cite{landau-lifshitz,carpio08,cadelano12}
\begin{equation}
U_s = \frac{1}{2} C_{iklm}\varepsilon_{ik} \varepsilon_{lm} -\beta_{ik}\Theta(x,y) \varepsilon_{ik} \delta_{ik},~~~~\label{U-s}
\end{equation}
where $C_{iklm}$ is a tensor of rank four  (the elastic modulus tensor),  $\varepsilon_{ik}$ (or $\varepsilon_{lm}$) is  the strain tensor and  $\beta_{_{ik}}$ are the  stress temperature coefficients. Also,  $\Theta(x,y)$ is the distribution of temperature in the graphene sheet that can be found by solving Laplace's equation $\partial_i q_i=0$, where $q_i=-\alpha_{_{ii}} \partial_{x_i} \Theta$ with $\alpha_{_{ii}}$ being  the heat induction  coefficients of graphene. Thus we write the Laplace's equation as
\begin{equation}
\alpha_{_{11}} \partial_x^2\Theta +\alpha_{_{22}} \partial_y^2\Theta=0.\label{kappa}
\end{equation}
We  suppose that the graphene sheet at the boundary 4 (see Fig.~\ref{fig1}) is connected to the heat bath of temperature $T(x)$  and all  other three boundaries 1,2,3  are fixed at zero temperature.  Thus, the exact solution of Laplace's equation can be written as
\begin{equation}
\Theta\left(x,y\right)=\sum_{m=1}^{\infty} B_m \sinh\left(\frac{m\pi y}{L\sqrt{k_e}}\right)\sin\left(\frac{m\pi}{L}x\right),\label{Theta}
\end{equation}
where the constant $B_m$  relates to the temperature of the thermal bath $T(x)$  as:
\begin{equation}
T(x)=\sum_{m=1}^{\infty} B_m \sinh\left(\frac{m\pi w}{L\sqrt{k_e}}\right)\sin\left(\frac{m\pi}{L}x\right).\label{Tx}
\end{equation}
Here $k_e=\alpha_{_{22}}/\alpha_{_{11}}$, m is an integer,  $L$ is the length and $w$ is the width of the graphene sheet. The arbitrary constant $B_m$ can be found by performing Fourier's transform of~(\ref{Tx}):
\begin{equation}
B_m=\frac{2}{ L \sinh\left(m\pi w/L\sqrt(k_e)\right)}\int_0^L T(x) \sin\left(\frac{m\pi x}{L}\right)dx.\label{Bm}
\end{equation}
In~(\ref{U-s}), the  strain tensor components can be written as
\begin{equation}
\varepsilon_{ik}=\frac{1}{2}\left(\partial_{x_k} u_i+\partial_{x_i} u_k  + \partial_{x_k} h \partial_{x_i} h   \right),\label{varepsilon-ik1}
\end{equation}
where $u_i$ and $h$ are in-plane and out-of-plane displacements, respectively.~\cite{juan13,carpio08}
It is believed that the ripples in graphene have two different types of sources (see Figs. 1 and 3 of Ref.~\onlinecite{thompson09}). Both types  of ripple waves  are considered to be the sinusoidal function of position whose amplitude lies normal to the plane of two-dimensional graphene sheet. The origin of first type ripples results in  the relaxed shape graphene (i.e., the displacement vector relaxed to the equilibrium position where the total elastic energy density is minimized) due to externally applied  in-plane (along x and y-directions only) tensile edge stress.~\cite{shenoy08} Such kind of ripples that can be seen in the form of relaxed shape graphene  occurs in a fashion similar to leaves or torn plastic~\cite{sharon02} where buckling mechanism displaces only the carbon atoms near the edge of the graphene sheet.~\cite{thompson09,shenoy08} While the origin of second type ripples results in  the height fluctuations through out the graphene sheet due to  adsorbed hydroxide molecules sitting on random sites of hexagon graphene molecules in a two-dimensional graphene sheet.~\cite{echtermeyer07,fasolino07,moser08}
In this paper, we only consider the ripple waves induced by buckling mechanisms ($i.e., h=0$) by applying tensile edge stress along the armchair and zigzag directions.~\cite{shenoy08,note}
Hence the strain tensor components for graphene in 2D displacement vector $\mathbf{u}(x,y)=(u_x,u_y)$ can be written as
\begin{eqnarray}
\varepsilon_{xx}=\partial_x u_x,~\varepsilon_{yy}=\partial_y u_y,~\varepsilon_{xy}=\frac{1}{2}\left(\partial_{y} u_x+\partial_{x} u_y\right) .\label{varepsilon-ik2}
\end{eqnarray}
The stress tensor components $\sigma_{ik}=\partial U_s / \partial \varepsilon_{ik}$ for graphene can be written as~\cite{zhou08}
\begin{eqnarray}
\sigma_{xx}=C_{11}\varepsilon_{xx}+C_{12}\varepsilon_{yy}- \beta_{11}\Theta,  \label{sigma-xx}\\
\sigma_{yy}=C_{12}\varepsilon_{xx}+C_{22}\varepsilon_{yy} - \beta_{22}\Theta ,  \label{sigma-yy}\\
\sigma_{xy}=2C_{66}\varepsilon_{xy}.  \label{sigma-xy}
\end{eqnarray}
In the continuum limit, elastic deformations of graphene sheets are described by the Navier equations $\partial_j \sigma_{ik}=0$. Hence, the  coupled Navier-type equations of thermoelasticity  for graphene can be written as
\begin{eqnarray}
\left(C_{11}\partial^2_x+C_{66}\partial^2_y\right)u_x  + \left(C_{12}+C_{66}\right) \partial_x \partial_y u_y =\beta_{11}\Theta_x,~~~~~   \label{coupled-1}\\
\left(C_{66}\partial^2_x+C_{22}\partial^2_y\right)u_y  + \left(C_{12}+C_{66}\right) \partial_x \partial_y u_x =\beta_{22}\Theta_y,~~~~~  \label{coupled-2}
\end{eqnarray}
where
\begin{eqnarray}
\Theta_x=   \sum_{m=1}^{\infty} \frac{m\pi B_m}{L} \sinh\left(\frac{m\pi y}{L\sqrt{k_e}}\right)\cos\left(\frac{m\pi}{L}x\right),\label{coupled-3}\\
\Theta_y= \sum_{m=1}^{\infty} \frac{m\pi B_m}{L\sqrt(k_e)} \cosh\left(\frac{m\pi y}{L\sqrt{k_e}}\right)\sin\left(-\frac{m\pi}{L}x\right). \label{coupled-4}
\end{eqnarray}
Now we apply the compressive tensile edge stresses  through the boundaries of the graphene sheet (see Fig.~\ref{fig1}) along the lateral directions and seek to establish the relationship between the waves generated due to  applied tensile edge stresses and the in-plane displacement vector.
More precisely, the mechanical deformation along the lateral   direction of a  graphene
sheet can be understood by considering the effect of externally applied  tensile edge
stresses  along the lateral direction in the form of compressed elastic string between which the 2-dimensional free standing  graphene sheet is
clamped.~\cite{landau_book,zhou08,cadelano09} It is natural to expect that the longitudinal waves, generated by compressed elastic forces, travel along the direction of applied tensile edge stresses.
Hence we might think  the sinusoidal function in the deformation of in-plane displacement vectors   that correlate  the geometric features of the longitudinal waves generated in the graphene sheet due to  compressed elastic tensile edge stresses  which finally turn   the flat surface of the graphene sheet into the relaxed shape graphene  in the form of torn plastic.
Thus  we assume the functional form of $\mathbf{u}\left(x,y\right)$ at the boundary $2$ of armchair as:
\begin{equation}
u(x,0)=u_x=A\sin\left(kx\right),\label{u}
\end{equation}
where $A$ is the amplitude and $k=2\pi/\lambda$ with $\lambda$ being the wavelength while  the total edge energy per unit length becomes~\cite{shenoy08}
\begin{equation}
U_e = \frac{1}{2} \tau_{e}\left(\frac{\partial u \left(x,0\right)}{\partial x}\right)^2 + \frac{1}{8} E_{e}\left(\frac{\partial u \left(x,0\right)}{\partial x}\right)^4,\label{U-e}
\end{equation}
where $\tau_{e}$ and $E_{e}$ denote the edge stress and the elastic modulus of the edge  along the armchair direction of the graphene sheet. Thus, we write the total edge energy  $U_t=\int_0^L U_e dx$ and find the optimum amplitude of the  waves along the armchair direction   by utilizing the condition $\partial U_t/\partial A=0$ to be:
\begin{equation}
A= \sqrt{ \frac{2\tau_e \left(2kL+\sin\left(2kL\right)\right)}{E_ek^2  \left[12kL+8\sin\left(2kL\right)+\sin\left(4kL\right)\right]}  }.  \label{A}
\end{equation}
By considering $L=1.5~\mathrm{\mu m}$, $\tau_e=4~\mathrm{eV/nm}$, $E_e=1000~\mathrm{eV/nm}$ and  $\lambda=0.1$ to $1.5~\mathrm{\mu m}$, the amplitude ($\mathrm{A}$) of the waves varies from $0.6~\mathrm{nm}$ to $8.7~\mathrm{nm}$.
This amplitude  decreases exponentially as it moves into the graphene sheet along y-direction. Thus we assume $u\left(x,y\right)=A\sin\left(kx\right)\exp{\left(-y/\ell\right)}$, where $\ell$ is the penetration depth of the ripple waves and write the total elastic energy of the graphene sheet as $\tilde{U}=U_t+\int^L_{x=0}\int^\infty_{y=0}U_s dx dy$. By  considering $\partial \tilde{U}/\partial A=0$, we found the functional form of the amplitude as
\begin{equation}
2A\Gamma_1+4A^3\Gamma_2-\Gamma_3+\Gamma_4=0, \label{cubic}
\end{equation}
where
\begin{widetext}
\begin{eqnarray}
\Gamma_1=\frac{k}{16}\left(2\tau_e+c_{11}\ell+c_{66}\ell\right)
\left\{\sin\left(2kL\right)+2kL\right\}
-\frac{1}{16k\ell}\left(c_{22}+c_{66}\right)
\left\{\sin\left(2kL\right)-2kL\right\}
-\frac{1}{4}\left(c_{12}+c_{66}\right)\sin^2\left(kL\right),\\
\Gamma_2=\frac{E_ek^3}{32}\left\{12kL+8\sin\left(2kL\right)+\sin\left(4kL\right)\right\},\\
\Gamma_3=\beta_{11}k\sum_{m=1}^{\infty} \frac{B_m}{2}\left\{ \frac{\cos\left(m\pi+kL\right)}{m\pi+kL}+\frac{\cos\left(m\pi-kL\right)}{m\pi+kL} -\frac{2m\pi}{m^2\pi^2-k^2L^2}\right\}\frac{m\pi L^2\ell^2 \sqrt{k_e}}{m^2\pi^2\ell^2-L^2\sqrt{k_e}},\\
\Gamma_4=\beta_{11}\sum_{m=1}^{\infty} \frac{B_m}{2}\left\{ \frac{\sin\left(m\pi+kL\right)}{m\pi+kL}+\frac{\sin\left(m\pi-kL\right)}{m\pi+kL} \right\}\frac{m\pi L^2\ell \sqrt{k_e}}{m^2\pi^2\ell^2-L^2\sqrt{k_e}}.
\end{eqnarray}
\end{widetext}
At absolute zero temperature (i.e., $ \Gamma_3=\Gamma_4=0$), the amplitude of the ripple waves obtained from Eq.~(\ref{cubic}) resembles the expression presented in Ref.~\onlinecite{shenoy08}.
Similar type of expression~(\ref{cubic}) can also be obtained for the optimum amplitude  of the longitudinal waves along the zigzag direction    by considering non-vanishing  $u\left(0,y\right)=u_y=A\sin{\left(ky\right)}$ in~(\ref{u}) and replacing $L$ by $w$ in~(\ref{A}).
\begin{figure}
\includegraphics[width=8.5cm,height=6cm]{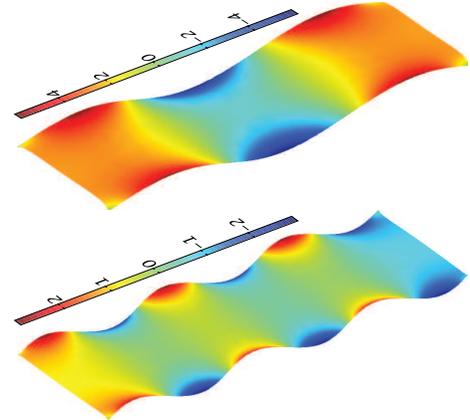}
\caption{\label{fig2}
Relaxed shape of graphene due to applied tensile stress along armchair  direction for  $\lambda=1~\mathrm{\mu m}$ (upper panel) and  $\lambda=0.5~\mathrm{\mu m}$ (lower panel).   Here we chose $T(x)=T_0=75~\mathrm{K}$, $\tau_e=4~\mathrm{eV/nm}$, $E_e=1000~\mathrm{eV/nm}$ and the dimension of the graphene sheet is taken to be  $L\times w = 1.5 \times 0.5~\mathrm{\mu m^2}$. The longer and shorter sides are considered as the armchair and zigzag edges respectively.
The variation in the amplitude of the ripple waves, expressed in nanometer, is shown in the color bar.
}
\end{figure}
\begin{figure}
\includegraphics[width=8.5cm,height=8cm]{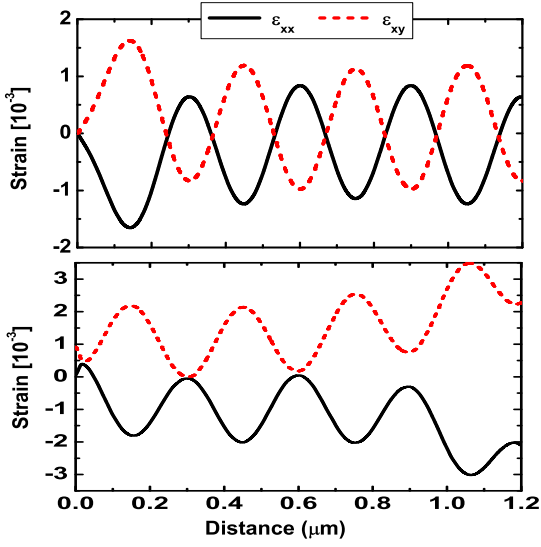}
\caption{\label{fig3}
Oscillations in strain tensor due to applied  tensile stress along the  armchair  direction at $y=w/2$ and temperature $T_0=0~\mathrm{K}$ (upper panel) and  $T_0=300~\mathrm{K}$ (lower panel). The parameters are chosen to be the same as in Fig.~\ref{fig2} but wavelength  $\lambda=300~\mathrm{nm}$.
}
\end{figure}
\begin{figure}
\includegraphics[width=8.5cm,height=6cm]{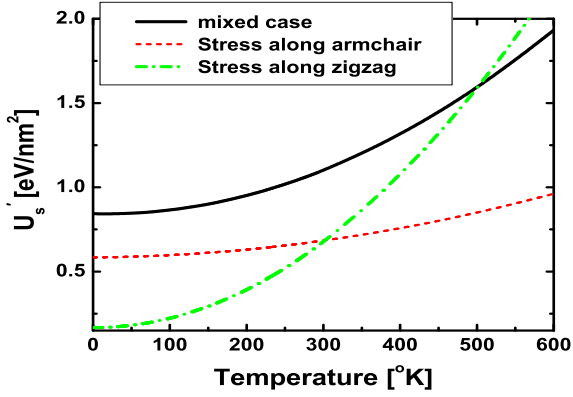}
\caption{\label{fig4}
Free elastic energy density, $U_s'=1/\tilde{A}\left(\int_\Omega U_s dx dy\right)$ with $\tilde{A}=L\times w$ and $\Omega$ being the computational domain. For mixed case, we have applied the tensile edge stress along both the armchair and zigzag direction.  The parameters are chosen to be the  same as in Fig.~\ref{fig2}.
}
\end{figure}
\begin{figure}
\includegraphics[width=8.5cm,height=6cm]{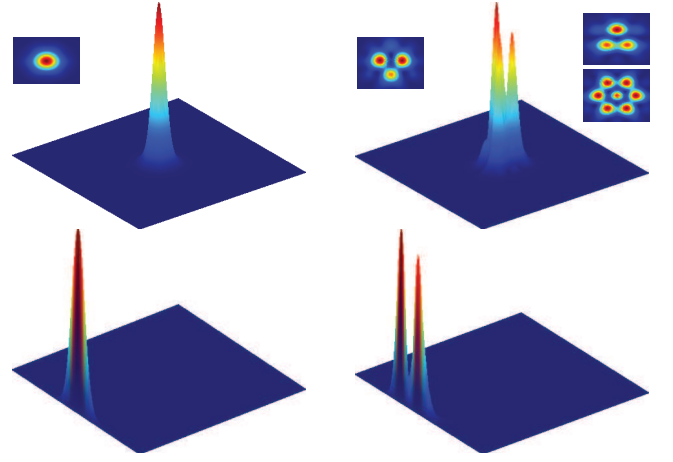}
\caption{\label{fig5}
Demonstration  of squared wavefunctions, $|\Psi|^2$ formed by lateral confined potential $U(x,y)$ (see Eq.~\ref{H}) in $L\times w=400\times 400~\mathrm{nm^2}$ graphene sheet. Upper panel shows the  states formed at the center of the graphene sheet and lower panel shows the edge states formed at the zigzag boundary 1 (see Fig.~\ref{fig1}). Note that in the upper panel,  unusual  first excited state wavefunction can be seen due to the presence of three fold symmetry in graphene. For eigenvalues see Fig.~\ref{fig6}. For electromechanical parts, we chose   $\tau_e=45\mathrm{eV/nm}$, $E_e=1000\mathrm{eV/nm}$ and  $T_0=5\mathrm{K}$ that mimic the experimentally reported values in Ref.~\onlinecite{lee08}. For bandstructure calculations, we chose  $U_0=0.2\mathrm{eV}$, $\lambda_1=\lambda_2=200\mathrm{nm}$ and dimensionless parameter $\xi=1.97$. 
}
\end{figure}
\begin{figure}
\includegraphics[width=8.5cm,height=6cm]{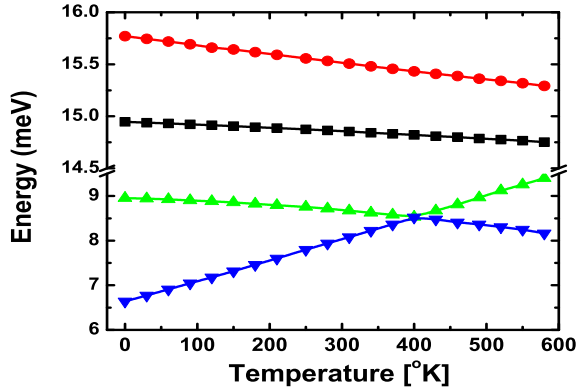}
\caption{\label{fig6}
Ground and first excited states eigenenergies  formed at the center of the graphene sheet (diamonds and circles) and at the edge of boundary 1 (triangles pointing up and triangles pointing down) vs temperature for the case $\tau_e=45~\mathrm{eV/nm}$ along the armchair direction. It can be seen that the level crossing between ground and first excited states edge energy occurs at $T=400~K$. The parameters are chosen to be the  same as in Fig.~\ref{fig5}.
}
\end{figure}
\begin{figure}
\includegraphics[width=8.5cm,height=6cm]{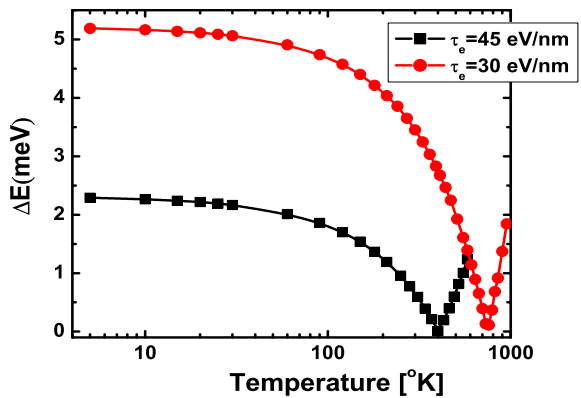}
\caption{\label{fig7}
Edge energy difference $\Delta E=E_1-E_2$ for the states at boundary 1 vs temperature. We can see that the level crossing point extends to larger temperatures with the decreasing values of the externally applied tensile edge stress along  the armchair direction.
}
\end{figure}

\begin{table}[b]
\caption{\label{table1}%
The material constants for graphene used in our calculations are taken from Refs.~\onlinecite{carpio08,shenoy08,shen10,duhee11,bao-lau09}.
}
\begin{ruledtabular}
\begin{tabular}{ll}
Parameters &  \\
\colrule
$C_{11}$[N/m]=$C_{22}$ & 359.4   \\
$C_{12}$[N/m] & 41  \\
$C_{66}$[N/m]  & 159.2 \\
$\alpha_{_{11}} [10^{-6}/\mathrm{K}]$  & -7 \\
$\alpha_{_{22}} [10^{-6}/\mathrm{K}]$  & -7  \\
$\beta_{11} [10^{-3}\mathrm{N}/\left(\mathrm{m} \cdot \mathrm{K}\right)]$$^a$   & -2.8 \\
$\beta_{22} [10^{-3}\mathrm{N}/\left( \mathrm{m} \cdot \mathrm{K}\right)]$$^a$   & -2.8 \\
\end{tabular}
\end{ruledtabular}
$^a$The numerical values of the thermal coefficients $\beta_{11}$ and $\beta_{22}$ are obtained from expressions $\beta_{11}=C_{11}\alpha_{_{11}}+C_{12}\alpha_{_{22}}$ and $\beta_{22}=C_{12}\alpha_{_{11}}+C_{22}\alpha_{_{22}}$.
\end{table}

Now we turn to the analysis of the influence of strain on the electronic properties of graphene QDs formed in the graphene sheet with the application of parabolic gate potential.
In the continuum limit, by expanding the momentum close to the $K$ point in the Brillouin zone, the Hamiltonian reads $H=\sum_k \Psi_k^\dagger \cdot H_k \cdot \Psi_k $. Here $H_k$ is written as~\cite{maksym13,krueckl12,neto09}
\begin{equation}
H_k =
\left(\begin{array}{cccc}
-U(x,y)&v{_{_F}} P_+& 0 & \frac{3\gamma_3 a P_-}{\hbar}\\
v{_{_F}} P_-& -U(x,y)& \gamma_1& 0\\
0&\gamma_1&U(x,y)& -v{_{_F}} P_-\\
\frac{3\gamma_3 a P_+}{\hbar} & 0  & -v{_{_F}} P_+ & U(x,y)
\end{array}\right),
\label{H}
\end{equation}
where $P_{\pm}=P_x\pm iP_y$ and $P=p-\hbar A$ with $p=-i\hbar \partial_x$ being the canonical momentum operator and $\textbf{A}=\beta \left(\varepsilon_{xx}-\varepsilon_{yy},\varepsilon_{xy}\right)/a$ is the vector potential induced by pseudomorphic strain tensor.~\cite{guinea08,juan13} Also,  $a$ is the lattice constant, $v_F=10^6\mathrm{m/s}$ is the Fermi velocity, $\gamma_3=0.3\mathrm{eV}$ corresponds to the interaction energy between two neighboring atoms A and B placed one under the other (see Ref.~\onlinecite{neto09})  and $\beta=-\partial \ln t/\partial \ln a \approx 2$ with $t$ being the nearest neighbor hoping parameters. We assume a confining potential $U(x,y)=U_x+U_y$ that decays exponentially away from the edges into the bulk with a penetration depths $\lambda_1$ or $\lambda_2$. Here we write $U(x)$ and $U(y)$ as:~\cite{neto09,maksym13}
\begin{eqnarray}
U(x)=U_0\left\{\exp{\left(\frac{-2x+L}{2\lambda_1}\right)} +\exp{\left(\frac{2x-L}{2\lambda_1}\right)} -\xi  \right\},~~~~~\label{Ux}\\
U(y)=U_0\left\{\exp{\left(\frac{-2y+w}{2\lambda_2}\right)} +\exp{\left(\frac{2y+w}{2\lambda_2}\right)} -\xi  \right\},~~~~~\label{Uy}
\end{eqnarray}
where $\lambda_1=L/2$, $\lambda_2=w/2$ and $\xi$ is  a dimensionless constant. We can  vary $\xi$ to vary the band gap of graphene induced by gate potential.

\section{Results and Discussions}\label{results-discussions}

The schematic diagram of the two-dimensional  graphene sheet in computational domain is shown in Fig.~\ref{fig1}. We have applied the tensile edge stress along both the armchair and zigzag directions to create  oscillations in  strain tensor of the graphene sheet. We have used the  multiphysics simulation and solved the Navier's  equations~(\ref{coupled-1}) and~(\ref{coupled-2}) via   Finite Element Method  to investigate the influence of thermo-electromechanical effects  on the relaxed shape of graphene. For the  waves along the armchair direction, we have used the Neumann boundary conditions at sides 1, 3 and employed Eq.~(\ref{u}) at sides 2,4 and vice versa for the zigzag direction. All reported results (Figs.~\ref{fig2}-\ref{fig4}) have been obtained for a  $1.5 \times 0.5 ~\mathrm{\mu m^2}$ graphene sheet that mimics the geometry of experimentally studied structures in Refs.~\onlinecite{bao-lau09}. All parameters for our simulations have been taken from Table~\ref{table1}. For bandstructure calculations (Figs.~\ref{fig5}-\ref{fig7}), we have chosen a $400 \times 400 ~\mathrm{nm^2}$ graphene sheet.

Fig.~\ref{fig2} shows the relaxed shape of graphene under applied tensile stress along the armchair direction.
The variation in the amplitude of  ripple waves, expressed in nanometer, is shown in the color bar (see Eq.~\ref{A}). As can be seen in the color bar, the amplitude of the ripple waves in the relaxed shape graphene is enhanced with the increasing values  of the wavelength which is also supported analytically by  Eq.~(\ref{A}).
In Fig.~\ref{fig3}, we investigate the influence of temperature on the  strain tensor under an applied tensile edge stress along the armchair direction. Again, the oscillations in the strain tensor  occur due to the  applied tensile stress along the armchair direction that induces longitudinal waves and  propagate along the armchair direction. We note that the increasing temperature from $0~\mathrm{K}$   (upper panel) to room temperature (lower panel) enhances the amplitude of the  waves that eventually increase the magnetic field of the pseudomorphic vector potentials and allow us to investigate its influence in the bandstructure calculation of graphene at Dirac points (for details, see Figs.~\ref{fig6} and~\ref{fig7}).
In Fig.~\ref{fig4}, we investigate the total elastic energy density vs temperature. Even though the optimum values of the amplitudes of  waves along the armchair and zigzag directions are exactly the same,  the variation in the  total  free elastic energy density  is enhanced for the case of  applied tensile edge stress along the zigzag direction (dashed-dotted line). This occurs because the graphene sheet along the boundary 4 is connected to the heat reservoir that enhances the free elastic energy of the  waves traveling along the zigzag direction.

Another important result of the paper is the study of  the influence of thermomechanics  on the bandstructure of bilayer graphene QDs via pseudomorphic vector potentials.
In Fig.~\ref{fig5}, we have plotted several states  wavefunctions of the graphene QDs. It can be seen that in addition to the localized states formed at the center of the graphene sheet, edge states are also present at the zigzag boundary.
The localized edge states at the zigzag boundary can be seen due to the fact that  the pseudomorphic vector potential originating from the thermo-electromechanical effects generates a large magnetic field applied perpendicular to the two-dimensional graphene sheet.
For example, considering $\mathbf{B}=\nabla \times \mathbf{A}$, we find $B_z=\left(2\pi\beta\varphi_0/a\right)\left\{\partial_x \varepsilon_{xy} -\partial_y \left(\varepsilon_{xx}-\varepsilon_{yy}\right)\right\}$, where $\varphi_0=\hbar/e$. Further, by assuming sinusoidal functions of  strain tensor (see Fig.~\ref{fig3})  generated  by applying tensile edge stress along the armchair direction, for example: $\varepsilon_{xx}=-\varepsilon_{xy}=\varepsilon_0\sin\left(kx\right)$ and $\varepsilon_{yy}=0$, where $\varepsilon_0$ is the amplitude of the strain tensor,   we find $B_z=-B_0\cos\left(kx\right)$, where $B_0=2\pi\varepsilon_0\beta k \hbar/a e$. By considering $\lambda=300\mathrm{nm}$ and $\varepsilon_0=10^{-3}$ (see Fig.~\ref{fig3} (upper panel)), we estimate $B_0\approx 1.2~ \mathrm{tesla}$.
Hence such a large  magnetic field along z-direction originating from electromechanical effects induces a persistent current that flows toward  the edge of the graphene. As a result, a positive  (negative) dispersion in the electron (hole) like states is induced at the zigzag edge (also see Fig.4 of Ref.~\onlinecite{neto06}) and the   localized electron-hole states wavefunction drift towards and over the edge of graphene sheet.~\cite{wen92,neto06,vikstrom14} Hence the  wavefunctions that are pressed against the edge  finally turn into the localized edge current carrying states.~\cite{neto06,zarenia11,vikstrom14,wen92}
Similar kind of localized edge states are also shown in Fig.~2(d) of Ref.~\onlinecite{neto06}. Furthermore, by utilizing tight binding and continuum model, the authors in Ref.~\onlinecite{zarenia11} have shown  that the tunneling effect at the interface between the internal and external regions of the dots can be avoided and consequently the charge carriers are confined at the zigzag edge (for example see Eq.~(11) and left panel of Fig.~6 of Ref.~\onlinecite{zarenia11}).
Such localized edge states,  shown in the lower panel of Fig.~\ref{fig5} in this paper, are highly sensitive to the applied tensile edge stress  along the armchair and  zigzag directions. We observe three fold symmetry in the first excited state wavefunction (see inset plot of Fig.~\ref{fig5}) of graphene that is in agreement to the experimentally reported results (see Fig.(4) of  Ref.~\onlinecite{klimov12}) and as well as previously reported theoretical results (see Ref.~\onlinecite{moldovan13}). In Fig.~\ref{fig6}, we  find the  level crossing at $T=400K$  due to the fact that the edge energy difference between the ground and first excited states  decreases with increasing temperature. This level crossing point extends to higher temperatures with decreasing values of tensile edge stress (see Fig.~\ref{fig7}). We have analyzed why the level crossing point can be seen on the edge states, but cannot be seen on the localized states  formed at the center of the graphene sheet. The  reason is that the graphene sheet on boundary  4 is connected to the heat reservoir. Hence,  the energy difference between the ground and first excited states  decreases with increasing temperature. As a result,  we find the level crossing in the edge states to be  at the zigzag boundary with the accessible values of  temperatures.
The energy difference between the ground and first excited states  formed at the center of the graphene sheet also decreases with increasing temperature (see Fig.~\ref{fig6}). However, such energy states do not meet each other with any practically applicable values of the temperature due to absence of two fold symmetry in graphene. In fact, the induced pseudomorphic fields by the strain tensor  affect the graphene charge carriers and  produce a three fold symmetry in the wavefunction of two-dimensional graphene sheet (see inset plot of~\ref{fig5}).
The symmetry of the pseudomorphic fields is determined by the corresponding symmetry of the strain field. For example, a uniform pseudomorphic field requires a special strain field distorted with three fold symmetry.~\cite{klimov12,guinea10}

\section{Conclusion}\label{conclusion}

To conclude, we have developed a model which allows us to investigate the influence of temperature on the relaxed shape of  the graphene sheet as well as in the QDs that are formed in the two-dimensional graphene sheet with the application of gate potentials. We have shown that  the variation in the  total  free elastic energy density  is enhanced with temperature for the case of  applied tensile edge stress along the zigzag direction.  We have treated the strain, induced by an applied tensile edge stress along the armchair and zigzag directions,  as a  pseudomorphic vector potential and shown that the level crossing point between the ground and first excited edge states at the zigzag boundary  extends to higher temperatures with decreasing values of the tensile edge stress. Such kind of level crossing is absent in the
states formed at the center of the graphene sheet due to the presence of three fold symmetry.

\begin{acknowledgments}
This work has been supported by NSERC and CRC programs (Canada). The authors acknowledge the Shared Hierarchical Academic Research Computing Network (SHARCNET) community  and Dr. P. J.  Douglas Roberts for his assistance and technical support.
\end{acknowledgments}


\begin{thebibliography}{44}
\bibitem [{not()}]{note}%
  \BibitemOpen
  \href@noop {} {}\bibinfo {note} {Our model only valid for a case where ripple
  waves can be generated via externally applied tensile edge stress where the
  amplitude of the ripple waves that lies normal to the plane of
  two-dimensional graphene sheet and decay exponentially as it moves into the
  graphene sheet similar to Ref.[4] and Fig.(1) of Ref.[26]. For a case where
  the ripple waves can be observed throughout the sample of graphene sheet
  similar to Ref.[27] of Fig.2, we can not assume $h=0$ in Eq.(6) and thus
  substiantial improvement of the developed model is necessary for such
  sineario.}\BibitemShut {Stop}%
\end{thebibliography}

%

\end{document}